# Cosmological Acceleration from Scalar Field and Classical and Quantum Gravitational Waves (Inflation and Dark Energy)

Leonid Marochnik, Physics Department, East-West Space Science Center, University of Maryland, College Park, MD 20742

We show that on the average, homogeneous and isotropic scalar field and on the average homogeneous and isotropic ensembles of classical and quantum gravitational waves generate the de Sitter expansion of the empty (with no matter) space-time. At the start and by the end of its cosmological evolution the Universe is empty. The contemporary Universe is about 70% empty, so the effect of cosmological acceleration should be very noticeable. One can assume that it manifests itself as dark energy. At the start of the cosmological evolution, before the first matter was born, the Universe is also empty. The cosmological acceleration of such an empty space-time can manifests itself as inflation. To get the de Sitter accelerated expansion of the empty space-time under influence of scalar fields and classical and quantum gravitational waves, one needs to make a *mandatory* Wick rotation, i.e. one needs to make a transition to the Euclidean space of imaginary time. One can assume that the very existence of inflation and dark energy could be considered as a possible observable evidence of the fact that time by its nature could be a complex value which manifests itself precisely at the start and by the end of the evolution of the Universe, i.e. in those periods when the Universe is empty (or nearly empty).

## 1. Introduction

Inflation and dark energy are two unsolved puzzles of modern cosmology. The idea of the necessity of inflation (exponentially fast expansion of the very early Universe) does not yet have direct reliable observational confirmation but it seems very attractive due to its ability to solve three known major cosmological paradoxes (flatness, horizon and monopoles) [1-3]. On the other hand, the existence of dark energy (exponentially fast expansion of the modern very late Universe) is an established observational fact [4, 5]. In case of inflation, it is almost generally accepted that the acceleration of expansion is due to a hypothetical scalar field. Choosing a different form of the potential of this field, the authors attempt to reconcile the theory with a number of e-foldings needed to ensure the flatness of the modern Universe to solve the horizon and monopole problems (references to the original works can be found in Weinberg's book [6]) and get an agreement with the CMB observations. In the case of dark energy, for the lack of a better choice, the cosmological constant and the quintessence ( evolving scalar field [7,8]) are considered as the popular candidates to provide an acceleration, although both ideas meet insuperable difficulties[1]. A common feature of both effects is exponentially rapid expansion of the

---

[1] One of these is a so-called "old cosmological constant problem": Why is the $\Lambda$-value measured from observations is of the order of $10^{-122}$ vacuum energy density? The second one is a "coincidence problem": Why is the acceleration happening during the contemporary epoch of matter domination?

Universe, and this is an accepted fact. Another common feature is the fact that both effects occur in an empty (or nearly empty) Universe. If an empty space should expand with acceleration then such a mechanism of acceleration should be common for both very early and very late Universe. The crucial importance of this fact was mentioned in [9]. The modern Universe is about 70% empty, so that we are in the emptying expanding Universe, which asymptotically must become completely empty, i.e. contain no matter. As a result of such emptiness, we observe the dark energy effect [9]. If before the birth of matter the Universe was empty, this empty space-time was to expand in accordance with the de Sitter law. It could explain the origin of inflation. The space-time without matter is not really empty as it always has the natural quantum and classical fluctuations of its geometry, i.e. gravitons and classical gravitational waves. The question arises: Could not gravitons and/or classical gravitational waves, filling the empty (or nearly empty) Universe, lead to its accelerated expansion? We have shown that the answer to this question is yes [10, 11]. As is shown in [9], this scenario is consistent with observational data. In this paper we show that in the FLRW metric the homogeneous and isotropic on the average scalar field with the constant potential produces the same self-consistent de Sitter state of the empty Universe as gravitons and classical gravitational waves do. In all these cases, it is necessary to use Wick rotation to get de Sitter solution to the equations of virtual gravitons, classical gravitational waves and scalar filed. In other words, to get the exponentially fast expansion of the empty FLRW space-time we must use time as a complex variable. The transition from Lorentzian to Euclidean metric indicates that this solution is of the instanton type.

## 2. Scalar Field

As is known (see, e.g. [12]), the energy density $\rho$ and pressure $p$ of a single real scalar field $\varphi(\vec{x},t)$ in the FLRW metric read

$$\rho = \frac{\dot{\varphi}^2}{2} + \frac{(\nabla \varphi)^2}{2a^2} + V(\varphi)$$
$$p = \frac{\dot{\varphi}^2}{2} - \frac{(\nabla \varphi)^2}{6a^2} - V(\varphi) \qquad (1)$$

Where $V(\varphi)$ is a potential of the scalar field, dots are time derivatives, $a(t)$ is a scale factor of FLRW metric, dots are time derivatives. The equation of motion for this field reads

$$\ddot{\varphi} + 3H\dot{\varphi} - \frac{1}{a^2}\nabla^2\varphi - \frac{\partial V}{\partial \varphi} = 0 \qquad (2)$$

Where $H = \dot{a}/a$ is the Hubble function. It is generally accepted in modern inflation theories employing scalar fields that terms containing $a^{-2}$ are negligible because of fast expansion and can be omitted (see e.g. [12]). For the descriptions of "old", "new", "eternal" and "chaotic" models of inflation see e.g. [13, 14, 15] and references therein. All these theories operate in the frame of the assumption that the terms containing $a^{-2}$ are negligible. In order for terms containing $a^{-2}$ not to interfere with the slow-roll conditions used in all modern models of inflation one needs to have large enough uniform patches $L$ which must be greater than $|\varphi/V'(\varphi)|^{1/2}$ [6]. In the frame of "chaotic" model of inflation [16], these patches can be chaotically distributed in the space-time. In a sense, in this work we also consider a chaotic scalar field but which is isotropic and homogeneous on average with the constant potential $V(\varphi) = const$. In such a case, the $a^{-2}$ terms are non-negligible and the slow-roll condition $L \gg |\varphi/V'(\varphi)|^{1/2}$ is not satisfied.

Using the Fourier transformation and averaging over 3-space, we get for (1)

$$\rho(t) =< \rho(\vec{x},t) >= \frac{1}{2}\sum_k <\dot{\varphi}_k \dot{\varphi}_k^* + \frac{k^2}{a^2}\varphi_k \varphi_k^* > +V$$
$$p(t) =< \rho(\vec{x},t) >= \frac{1}{2}\sum_k <\dot{\varphi}_k \dot{\varphi}_k^* - \frac{k^2}{3a^2}\varphi_k \varphi^*_k > -V \quad (3)$$

Where superscript ($*$) is the sign of complex conjugation. The transition from summation to integration, taking into account the isotropy of space can be done in the following way

$$\sum_{\mathbf{k}} \ldots \to \int d^3k/(2\pi)^3 \ldots = \int_0^\infty k^2 dk/2\pi^2 \ldots . \quad (4)$$

It is also convenient to make the transition from physical time $t$ to the cosmological time $\eta = \int dt/a$. Thus, Eqns. (3) and (2) may be rewritten in the following form

$$\rho =< \rho(\vec{x},\eta) >= \frac{1}{4\pi^2 a^2}\int_0^\infty k^2 dk <\varphi'_k \varphi'^*_k + k^2 \varphi_k \varphi^*_k > +V \quad (5)$$

$$p =< p(\vec{x},\eta) >== \frac{1}{4\pi^2 a^2}\int_0^\infty k^2 dk <\varphi'_k \varphi'^*_k - \frac{k^2}{3} \varphi_k \varphi_k^* > -V \quad (6)$$

$$\phi_k'' + (k^2 - \frac{a''}{a})\phi_k = 0 \qquad \varphi_\mathbf{k} = \frac{1}{a}\phi_\mathbf{k} \quad (7)$$

Where primes are derivatives over cosmological time $\eta$. To get a self-consistent description of the system, one needs to add Einstein's equations to the set of equations (5-7)

$$3\frac{a'^2}{a^4} = \kappa\rho \tag{8}$$

$$2\frac{a''}{a^2} - \frac{a'^2}{a^4} = -\kappa p \tag{9}$$

with $\kappa = 8\pi G$. In conformal time, the de Sitter expansion reads

$$a = -(H\eta)^{-1} \tag{10}$$

Below we show that (10) is the exact solution to the set of equations (5-9). The solution to (7) over the de Sitter background (10) reads (see, e.g. [17, 18, and 10])

$$\varphi_k(\eta) = \frac{1}{a\sqrt{2k}}[Q_k(1-\frac{1}{ik\eta})e^{-ik\eta} + Q^*_k(1+\frac{1}{ik\eta})e^{ik\eta}] \tag{11}$$

Here $Q_k$ is the integration constant. The solution (11) can be rewritten also in the following useful form

$$\varphi_k(x) = H(2k^{-3})^{1/2}[A_k \cdot (x\cos x - \sin x) + B_k \cdot (x\sin x + \cos x)], \quad x = k\eta \tag{12}$$

Here $A_k = \operatorname{Re} Q_k$ and $B_k = \operatorname{Im} Q_k$.

To make sure that (10) is an exact solution to the set of eqns. (5-7), we have to substitute (10) and (11) into (5-6) and find $H$ for which this set of equations is satisfied.

The substitution of (11) into Eqns. (5-6) leads to divergent integrals. To calculate integrals in (5-6), we make Wick rotation $\eta = -i\varsigma$, $\xi = k\varsigma$ in Eqns. (5), (7) and (10). Instead of (5), (7) and (10) we get the following

$$\rho = <\rho(\vec{x},\eta)> = \frac{1}{4\pi^2 a^2}\int_0^\infty k^2 dk < -\varphi'_k\varphi'^*_k + k^2\varphi_k\varphi^*_k > +V \tag{13}$$

$$\phi''_k - (k^2 + \frac{a''}{a})\phi_k = 0 \quad \varphi_{\mathbf{k}} = \frac{1}{a}\phi_{\mathbf{k}} \tag{14}$$

$$a = 1/iH\varsigma \tag{15}$$

Primes in Eqns. (13-14) indicate derivatives over $\varsigma$.

Instead of (11), we get the following solution for (14) over the background (15)

$$\varphi_k(\xi) = -iH(2k^3)^{-1/2}[q_k(\xi+1)e^{-\xi} + q_k^*(\xi-1)e^{\xi}] \tag{16}$$

To get a finite solution, one has to choose $q_k^* = a_k - ib_k = 0$. Thus, we get

$$\varphi_k(\xi) = H(2k^{-3})^{1/2} \cdot b_k \cdot (\xi+1)e^{-\xi} \qquad (17)$$

The substitution of (17) to (13) leads to the following equation for the energy density

$$\rho = \frac{H^4}{2\pi^2}\int_0^\infty <b_k^2>[\xi^2-(1+\xi)^2]e^{-2\xi}\xi d\xi + V \qquad (18)$$

For the flat spectrum $<b_k^2>=<b^2>=const$, the integral in Eqn. (18) reads

$$\int_0^\infty [\xi^2-(1+\xi)^2]e^{-2\xi}\xi d\xi = -3/4$$

From (18) we get

$$\rho = -\frac{3<b^2>H^4}{8\pi^2} + V \qquad (19)$$

In accordance with (15), after Wick rotation the LHS of (8) is $-3H^2$. From (8), (15) and (19) we get the following equation for the Hubble constant $H$

$$-3H^2 = -\frac{3\kappa<b^2>H^4}{8\pi^2} + \kappa V \qquad (20)$$

Eqn. (7) is a Schrödinger-like equation with the "one-dimensional potential" $a''/a$. The only a difference is that the spatial coordinate in the Schrödinger equation is changed for the time variable $\eta$ in Eqn. (7). In such terms, the solution (12) is a superposition of incident and reflected waves with respect to the "barrier" $x=\xi=0$, and solution (17) is a transition wave. To make the analytical continuation from (17) back to real time (reverse Wick rotation), we have to satisfy two following conditions (see also [11])

$$\varphi_k(\xi=0) = \varphi_k(x=0) \qquad (21)$$

$$\varphi'_k(\xi=0) = \varphi'_k(x=0) \qquad (22)$$

The (22) condition is satisfied automatically because of $\varphi'_k(\xi=0) = \varphi'_k(x=0) = 0$. The condition (21) is satisfied if $b_k = B_k$. We get

$$<b_k^2>=<b^2>=<B^2>$$

Finally, Eqn. (20) can be rewritten in the following form

$$3H^2 = \frac{3\kappa <B^2> H^4}{8\pi^2} - \kappa V \qquad (23)$$

In Eqn. (23), $<B^2>^{1/2}$ is the root mean square of amplitude of scalar field oscillations. In the case when $V = 0$, we get

$$H^2 = \frac{\pi}{G<B^2>} \qquad (24)$$

Thus, a chaotic scalar field which is homogenous and isotropic on average (with potential $V = const$) generates the de Sitter expansion of the empty space. The speed of such expansion is determined by (23) or (24). Note that we have here the vacuum equation of state

$$-p = \rho = \frac{3<B^2> H^4}{8\pi^2} - V \qquad (25)$$

## 3. Gravitational Waves

In the frame of the self-consistent approach, the state of gravitational waves (GW) is determined by their interaction with background geometry, and the background geometry, in turn, depends on the state of GW. In such an approach, the interaction between gravitational waves is considered negligible. As was already mentioned, the back reaction of the long gravitational waves was considered in a number of works ([19], [20] and references therein) where the authors used linear parameterization of the metric tensor $g^{ik} = g_0^{ik} + h^{ik}$, where $g_0^{ik}$ and $h^{ik}$ are background metric tensor and its perturbation, respectively. This parameterization leads to the non-conservative energy momentum tensor (EMT) for the long gravitational waves. The existence of this problem was exhaustively documented in [19]. The solution to the problem was given in [21] where it was shown that it is necessary to use exponential parameterization corresponding to the use of normal coordinates in the functional space. With this parameterization, the EMT of gravitational waves satisfies the conservative condition automatically, and the self-consistent wave and background equations must be ensured by the variational principle ([21], section II.D). The equations describing this work can be obtained directly from the variational principle (see Eqs. (II.35) and (II.36) of [21])) or from the set of equations of one-loop quantum gravity (see sections II.D and III] of [21]) by the transition to the classical limit. They read

$$3H^2 = \kappa(\rho_g + \Lambda) \qquad (26)$$

$$2\dot{H} + 3H^2 = \kappa(\Lambda - p_g) \qquad (27)$$

$$\rho_g = \frac{1}{8\kappa} \sum_{\mathbf{k}\sigma} <\dot{\psi}^*_{\mathbf{k}\sigma}\dot{\psi}_{\mathbf{k}\sigma} + \frac{k^2}{a^2}\psi^*_{\mathbf{k}\sigma}\psi_{\mathbf{k}\sigma}> \qquad (28)$$

$$p_g = \frac{1}{8\kappa} \sum_{\mathbf{k}\sigma} <\dot{\psi}^*_{\mathbf{k}\sigma}\dot{\psi}_{\mathbf{k}\sigma} - \frac{k^2}{3a^2}\psi^*_{\mathbf{k}\sigma}\psi_{\mathbf{k}\sigma}> \qquad (29)$$

Where $\Lambda$ is the cosmological constant; Eqns. (26) and (27) are Einstein's equations for the background and (28, 29) form the EMT of GW; $\rho_g$ and $p_g$ are the energy density and pressure of GW; $\sigma$ is the polarization index; $\psi_{\mathbf{k}\sigma}$ is the Fourier image of tensor fluctuations (GW) that are gauge–invariant by definition. The mathematically rigorous method of separating the background and the GW, which ensures the existence of the EMT, is based on averaging over polarizations $<\psi^\beta_\alpha> \equiv 0$ of GW if all polarizations are equivalent in the homogeneous isotropic GW ensemble [21]. About the gage invariance problem and the elimination of 3–scalar and 3–vector modes see [21, section III.A]. As is known, the equation for 3–tensor GW is

$$\psi^\beta_{\alpha(t)}(t,\mathbf{x}) = \sum_{\mathbf{k}\sigma} Q^\beta_\alpha(\mathbf{k}\sigma)\psi_{\mathbf{k}\sigma}(t)e^{i\mathbf{k}\mathbf{x}}, \qquad \ddot{\psi}_{\mathbf{k}\sigma} + 3H\dot{\psi}_{\mathbf{k}\sigma} + \frac{k^2}{a^2}\psi_{\mathbf{k}\sigma} = 0 \quad (5) \quad (30)$$

The transition from summation to integration in Eqns. (28, 29), taking into account the isotropy of space, can be done by method (4). After transition to the cosmological time $\eta$, Eqns. (28), (29) and (30) can be transformed into the following form

$$\rho_g = \frac{1}{16\pi^2 a^2}\int_0^\infty k^2 dk <\hat{\psi}'^*_{\mathbf{k}\sigma}\hat{\psi}'_{\mathbf{k}\sigma} + k^2\hat{\psi}^*_{\mathbf{k}\sigma}\hat{\psi}_{\mathbf{k}\sigma}>$$

$$p_g = \frac{1}{16\pi^2 a^2}\int_0^\infty k^2 dk <\hat{\psi}'^*_{\mathbf{k}\sigma}\hat{\psi}'_{\mathbf{k}\sigma} - \frac{k^2}{3}\hat{\psi}^*_{\mathbf{k}\sigma}\hat{\psi}_{\mathbf{k}\sigma}> \qquad (31)$$

$$\chi''_{\mathbf{k}\sigma} + (k^2 - \frac{a''}{a})\chi_{\mathbf{k}\sigma} = 0, \qquad \psi = \frac{1}{a}\chi \qquad (32)$$

It is easy to see that the set of Eqns. (26), (27), (31) and (32) for the self-consistent GW problem is the same as Eqns. (5-9) for the scalar field with a constant potential. The only difference is in notation, substitution V for $\Lambda$ and numerical coefficient in Eq. (5) which takes into account two polarizations of GW. For example, in case of $\Lambda = V = 0$, instead of (24) we get for GW the following[2]

$$H^2 = \frac{8\pi}{G<Q^2>} \qquad (33)$$

---

[2] In our work [11], there is a misprint in formulas (46-49). Instead of $Q^2$ must be $\kappa Q^2$

Here $<Q^2>^{1/2}$ is the root mean square of GW amplitude averaged over their polarizations. The backreaction of classical gravitational waves and the scalar field with a constant potential lead to the same consequence which is de Sitter accelerated expansion of the empty FLRW space-time. As well as in the case of a scalar field, to get the solution (33) we must make Wick rotation and carry out the integration along the imaginary time axis. And as well as in the case of scalar field, we have here the vacuum equation of state

$$-p = \rho = \frac{3<Q^2>H^4}{64\pi^2} \qquad (34)$$

## 4. Virtual gravitons

The quantum theory of gravitons and its one-loop approximation are presented in [10, 21]. The self-consistent equations of one-loop approximation for FLRW metric are (III.7-III.18)] from [21] that can be rewritten in the following form

$$3H^2 = \kappa(\rho_g + \Lambda) \qquad (35)$$

$$2\dot{H} + 3H^2 = \kappa(\Lambda - p_g) \qquad (36)$$

$$\rho_g = \frac{1}{16\pi^2 a^2} \int_0^\infty k^2 dk <\Psi | \hat{\psi}'^*_{\mathbf{k}\sigma}\hat{\psi}'_{\mathbf{k}\sigma} + k^2 \hat{\psi}^*_{\mathbf{k}\sigma}\hat{\psi}_{\mathbf{k}\sigma} | \Psi >_{ren}, \qquad (37)$$

$$p_g = \frac{1}{16\pi^2 a^2} \int_0^\infty k^2 dk <\Psi | \hat{\psi}'^*_{\mathbf{k}\sigma}\hat{\psi}'_{\mathbf{k}\sigma} - \frac{k^2}{3}\hat{\psi}^*_{\mathbf{k}\sigma}\hat{\psi}_{\mathbf{k}\sigma} | \Psi >_{ren}, \qquad (38)$$

Where $\Psi$ is state vector and subscript "ren" (renormalized) indicates that in the calculation of average values in the structure of the graviton vacuum the ghost sector is taken into account. The brackets in Eqns. (37-38) indicate averaging over the quantum graviton ensemble. Eqns. (35) and (36) are Einstein's equations for the FLRW background and (37, 38) form the EMT of gravitons. In these equations $\rho_g$ and $p_g$ are the energy density and pressure of gravitons; $\psi_{\mathbf{k}\sigma}$ is the Fourier image of tensor fluctuations that are gauge–invariant by definition. The mathematically rigorous method of separating the background and the gravitons, which ensures the existence of the EMT, is based on averaging over polarizations $\sigma$ $<\psi^\beta_\alpha> \equiv 0$ of gravitons if all polarizations are equivalent in the homogeneous isotropic graviton ensemble [21]. As is known, the equation for 3–tensor GW is

$$\psi^\beta_\alpha(t,\mathbf{x}) = \sum_{\mathbf{k}\sigma} Q^\beta_\alpha(\mathbf{k}\sigma)\psi_{\mathbf{k}\sigma}(t)e^{i\mathbf{k}\mathbf{x}}, \qquad \ddot{\psi}_{\mathbf{k}\sigma} + 3H\dot{\psi}_{\mathbf{k}\sigma} + \frac{k^2}{a^2}\psi_{\mathbf{k}\sigma} = 0 \qquad (39)$$

To Eqns. (35-39) must be added canonical commutation relations and Grassman ghost field $\hat{\theta}_k$ which is also described by Eqn. (39) [21]. In conformal time, Eqn. (39) coincides with (7) and (32). It is easy to see that the set of self-consistent equations of scalar field (5-9) coincides with the set of self-consistent equations for GW (26, 27, 31, and 32) and with the set of self-consistent equations for gravitons (35-39) up to notation, normalization factors, replacement $\Lambda \rightarrow V$ and the appearance of ghosts in the quantum case. As is shown in [22], repeating procedures (11-20) we get in this quantum case (40) instead of (23) and (25)

$$3H^2 = \frac{3\kappa \hbar N_g}{8\pi^2} H^4 + \Lambda \qquad (40)$$

Where $N_g$ is the graviton occupation number (renormalized by ghosts) which is close to the number of gravitons in the Universe[3]. It is important to note that Eqn. (40) also obtained by Wick rotation as well as Eqns. (23) and (33). For $\Lambda = 0$ we get similarly to (33)

$$H^2 = \frac{\pi}{G\hbar N_g} \qquad (41)$$

Despite the external similarity there is a significant difference between classical and quantum cases. The graviton-ghost system forms the coherent quantum condensate responsible for the macroscopic effects of quantum gravity, which is the de Sitter expansion [21]. The de Sitter solutions for isotropic homogeneous on average scalar field, GW and gravitons are obtained for the empty Universe, i.e. they are applicable to the beginning of the Universe evolution (before matter was born) and to the end of it (after the matter disappears). It can be assumed that they are applicable to the inflation and dark energy. The questions arise when the inflation stopped and when the dark energy appeared and became detectable? For gravitons, answers to these questions were given in works [22] and [9]. The same is applicable to the case of GW and scalar field. The inflation stopped when the mass of newly born matter became large enough to "spoil" a solution for the empty space. If we accept the estimations of e-foldings $\mathbb{N} \sim 62-68$ before the inflation stopped [6], then one can say that this is the time when the mass of the newly born matter is sufficient to stop the de Sitter expansion. Similar, but in a sense, the opposite situation occurs in the case of dark energy [22]. The dark energy appears when the energy density of existing matter becomes small enough not to spoil the de Sitter solution which originates in the emptying Universe. As was shown in our works [22, 9], the de Sitter expansion (41) leads to CMB fluctuations of the order of $\Delta T / T \approx 10^{-5}$, i.e. it is consistent with observational data. It also provides a natural explanation to the "coincidence problem" (why is the dark energy acceleration happening during contemporary epoch of matter domination?).

---

[3] Despite the widespread belief that the ghost's contribution should not appear in the final result of the calculations, in fact, this is only true for the asymptotic states as it takes place in the S-matrix theory. There are no asymptotic states in the Universe which as a whole is the region of interactions. This is the reason why we deal with virtual gravitons (strictly speaking, there are no real gravitons in the Universe). Ghosts appear here to compensate the effect of vacuum polarization of fictitious fields of inertia [22]. This is the reason why they appear only as a factor renormalizing occupation numbers of gravitons.

## 5. Wick rotation and time as a complex variable

A textbook case is the non-stationary Schrödinger equation which can be transformed into a heat or diffusion equation by Wick rotation $t = i\tau$. Solutions to these equations are well known. One gets a solution to the non-stationary Schrödinger equation by such a simple way. This fact gave rise to a widespread belief that the Wick rotation is simply a convenient change of variables. As a matter of fact, the situation with this rotation is not so simple. As is known, the imaginary time formalism is used in non–relativistic Quantum Mechanics (QM) (see, e.g., [23]), in the instanton theory of Quantum Chromodynamics (QCD) (see, e.g. [24]) and in the axiomatic quantum field theory (AQFT) (Chapter 9 in [25]). The instanton physics in Quantum Cosmology was discussed in [26, 27, 21] and references therein.

In QM and QCD the imaginary time formalism is a tool for the study of tunneling, uniting classic independent states that are degenerate in energy, into a single quantum state. In non–relativistic QM the barriers are considered that have been formed by classical force fields and for that reason they have the obvious interpretation. A new class of phenomena arises in the cases when tunneling processes form a macroscopic quantum state. The Josephson Effect is a characteristic example: fluctuations of the electromagnetic field arise when a superconductive condensate is tunneling across the classically impenetrable non–conducting barrier. Here, *the tunneling can be formally described as a process developing in imaginary time, but the fluctuations arise and exist in the real spacetime.* Experimental data show that regardless of the description, the tunneling process forms a physical subsystem in the real space–time with perfectly real EMT. In this work, we deal with similar situation in cosmology. Recall that the self-consistent sets of equations for scalar field (5-9), GW (31-32) and gravitons (35-39) have no solutions in real time because of divergences of integrals. The self-consistent solutions to these equations exist in imaginary time only[4]. Quantum metric fluctuations in imaginary time (graviton-ghost instantons) form a macroscopic de Sitter state in real time as do GW and scalar field. This fact is supposed to be confirmed by observational data which could be the existing dark energy and inflation effects. As we see, usually Wick rotation is used in the quantum instanton theories. However, as was shown in Sections 2 and 3, Wick rotation leads to the same de Sitter expansion of empty space in classical cases too, and this means that a central role is played not by the quantum nature but the transition to imaginary time by Wick rotation.

Recall that the integration along real time axis in (5) leading to divergent integrals should be replaced by integration along the imaginary time axis in (13) leading to convergent integrals. This means that we operate in the plane of complex variable $t$ (or $\eta$), i.e. we use time as a complex variable. Time as a complex variable is just a mathematical trick that leads to the desired result, or is there behind it certain new physics? The mathematical aspects of time as a complex quantity was extensively discussed in literature (see [28] and references therein).The best known attempt to ascribe physical meaning to the imaginary time was made by Hartle and Hawking [29]. Later, Hawking emphasized repeatedly in his books on popular science a possible reality of imaginary time (see, e.g. [30]). As is shown in this paper, the de Sitter accelerated expansion of the empty FLRW space under backreaction of quantum metric fluctuations, classical gravitational waves and/or scalar field require a *mandatory* transition to the Euclidean space of imaginary time and then return to the Lorentzian space of real time. On the other hand, the de Sitter accelerated expansion of the empty space at the beginning and end of the evolution of the Universe is confirmed by observational data (dark energy and inflation). One can assume that the very existence of these two effects is the observable evidence to the fact that time by its nature could be a complex value in the empty spacetime of the Universe[5]. If time is a complex variable, what is the physical meaning of its imaginary part? I am indebted to Daniel Usikov profound remark that the observable could be only the real part of time, and the imaginary part, for whatever reasons, is unobservable (by analogy with quantum mechanics).

---

[4] In the case of virtual gravitons the situation is not so unambiguous. For instance, de Sitter solution can be obtained in real time (without Wick rotation) but price for that is a "materialization" of ghosts which are non-physical particles [10, 21, and 22].

**Acknowledgment**

I am deeply grateful to Daniel Usikov for useful discussions. I would like to express my deep appreciation to Walter Sadowski for invaluable help in the preparation of the manuscript.

# References

1. A. Guth, Phys.Rev. D23, 347, 1981

2. A.D. Linde, Phys. Lett. B 108, 389, 1982

3. A. Albrecht & P.Steinhardt, Phys.Rev. Lett 48, 1220, 1982

4. A.G. Riess et al, AJ, **116**, 1009, 1998

5. S. Perlmutter et al., ApJ, **517**, 565, 1999

6. S. Weinberg, Cosmology, Oxford Univ. Press, 2008

7. P. Ratra & L.Peebles, Phys. Rev. D 37, 3406, 1988; Rev. Mod. Phys. 75, 559, 2003

8. R. Caldwell, R. Dave & P.J. Steinhardt, Phys. Rev.Lett. 80, 1582, 1998

9. L. Marochnik & D. Usikov, Grav. Cosmology 21, 118, 2015

10. L. Marochnik, D. Usikov & G. Vereshkov, Found.Phys.38, 546, 2008;

11. L.Marochnik, Grav. Cosmology 22, 10, 2016

---

[5] The fact that this scenario is consistent with observational data on inflation and dark energy [9] makes a speculative hypothesis about the complex nature of time less speculative than it seems at first glance.

12. E. Kolb & M. Turner, The Early Universe, 1990, Frontiers in Physics, Addison-Wesley

13. A. Guth, Pritzker Symp. on the Status of Inflationary Cosmology, 1999; astro/ph0002188

14. A. Linde, J. Phys. Conf. Ser. 24, 151, 2005

15. A.Linde, New Astronomy Reviews, 49, 35, 2005

16. A. Linde, Phys. Lett. 129B, 177, 1983

17. T. S. Bunch & P.C.W. Davies, Proc. Roy. Soc. London, A360, 117, 1978

18. A. Starobinsky, Pisma JETP, 719, 1979

19. L.R.W. Abramo, R. Brandenberger & V. Mukhanov, Phys. Rev. **D56**, 3248, 1997

20. A. Ishibashi & R.Wald, Class. Quantum Grav. **23 ,** 235–250, 2006.

21.  L. Marochnik, D. Usikov & G. Vereshkov, arXiv: 0811.4484, 2008; ibid JMP, 4, 48, 2013

22. L. Marochnik, Grav. Cosmology, 19, 178, 2013; arXiv: 1306.6172

23. R. Rajaraman, "Solitons and Instantons. An Introduction to Solitons and Instantons in Quantum Field Theory," North-Holland Publishing Company, Amsterdam, 1982

24. M. A. Shifman, ed. "Instantons in Gauge Theories," World Scientific, Singapore, 1994

25. N. N. Bogolyubov, A. A. Logunov, A. I. Oksak & I. T. Todorov, "General Principles of Quantum Field Theory," Kluwer, Springer, 1990

26. G.W. Gibbons & S.W. Hawking, Eds, "Euclidean Quantum Gravity", World Scientific, Singapore, 1993

27. V. A. Rubakov & P. G. Tinyakov, Physics Letters B, 214, 334, 1988

28. G. Esposito, Int. Journal of Geometric Methods in Modern Physics, pp.1-59, 2005; arXiv: hep-th/0504089v2

29. B.J. Hartle & S.W. Hawking, Phys. Rev, D28, 2960, 1983

30. S. Hawking, The Universe in Nutshell, Bantam Books, 2001